# SM-LIKE HIGGS DECAY INTO TWO MUONS AT 1.4 TEV CLIC

I. Bozovic-Jelisavcic*[1], S. Lukic*, G. Milutinovic-Dumbelovic*, M. Pandurovic*

*Vinca Institute of Nuclear Sciences, University of Belgrade, Serbia
[1] speaker

[on behalf of the CLIC Detector and Physics Collaboration]

**Abstract**

The branching fraction measurement of the SM-like Higgs boson decay into two muons at 1.4 TeV CLIC will be described in this paper contributed to the LCWS13. The study is performed in the fully simulated ILD detector concept for CLIC, taking into consideration all the relevant physics and the beam-induced backgrounds, as well as the instrumentation of the very forward region to tag the high-energy electrons. Higgs couplings are known to be sensitive to BSM physics and we prove that BR times the Higgs production cross section can be measured with approximately 35.5% statistical accuracy in four years of the CLIC operation at 1.4 TeV centre-of-mass energy with unpolarised beams. The result is preliminary as the equivalent photon approximation is not considered in the cross-section calculations. This study complements the Higgs physics program foreseen at CLIC.

# 1. Introduction

Measurement of the Higgs branching ratios and consequently Higgs couplings provide strong test of the Standard Model (SM) and the physics beyond. Models that could possibly extend SM Higgs sector (2DHM, Little Higgs models or Compositeness) will require Higgs couplings to electroweak bosons and Higgs Yukawa couplings (coupling-mass linearity) to deviate from the Standard Model predictions. Higgs sector measurements at a future linear collider can provide insight into EWSB mechanism as well as in physics beyond the Standard Model (BSM).

Compact Linear Collider (CLIC) represents an excellent environment to study properties of the Higgs boson, including Higgs couplings with a very high precision [1]. Measurement of the rare H→$\mu^+\mu^-$ decay is particularly challenging due to the very low branching ratio of the order of $10^{-4}$. The measurement thus requires excellent muon identification efficiency and momentum resolution as well as comprehensive background suppression.

This paper we describe the full analysis chain to measure Higgs branching fraction into two muons and to extract the corresponding coupling $g_{H\mu\mu}$. Presented results are preliminary since no Equivalent photon approximation (EPA) is considered in the cross-section calculations in the kinematical region where the momentum transferred by photon is small (i.e. $Q^2 < 4$ GeV$^2$).

This analysis is part of the detector benchmark studies at CLIC [7].

# 2. Simulation tools

Higgs production through $W^+W^-$ fusion is simulated in WHIZARD v 1.95 [2] including realistic beam spectrum and the initial state radiation. PYTHIA 6.4 [4] is used to simulate the Higgs decay into two muons, including the effect of final state radiation. Background events are also generated in WHIZARD using PYTHIA to simulate hadronisation and fragmentation processes. Tau decays are provided by TAUOLA [5]. The CLIC luminosity spectrum and the beam induced processes are obtained by GuineaPig 1.4.4 [6]. A Lorentz boost of the final state particles is performed in order to simulate 20 mrad beam crossing angle.

Interactions with the detector are simulated in CLIC_ILD detector model in Mokka [7,8]. A particle flow algorithm [9] is employed in reconstruction of the final state particles.

The TMVA package [10] is used in a multivariate analysis of signal and background kinematic properties.

# 3. CLIC_ILD detector model

The ILD detector model has been modified according to specific requirements for CLIC and we will discuss here only the subsystems performances of particular relevance for this analysis. A detailed description of the CLIC_ILD detector can be found in [7].

High efficiency muon identification is provided by the iron yoke, which is instrumented with 10 layers of RPC detectors. Information from RPCs is then linked to the inner tracker in order to obtain muon track candidates. Efficiency of muon identification is influenced by the iron longitudinal segmentation as well as by the background processes. Hadrons produced in the interaction of beamstrahlung photons can increase occupancy of the inner tracker and consequently deteriorate muon identification efficiency. On the signal sample of muons from H→$\mu^+\mu^-$ decay, muon efficiency is above 99% in the barrel region. This analysis is particularly challenging to detector momentum resolution as it influences the width of the reconstructed di-muon invariant mass. In the present analysis, the average muon transverse momentum resolution on the signal sample in the barrel region is $\Delta p_T / p_T^2 = 3.3 \times 10^{-5} \, \text{GeV}^{-1}$.

The forward region below 6.9 deg of the CLIC_ILD detector is instrumented with two silicon-tungsten sampling calorimeters, a luminometer and a beam-calorimeter that provide high-energy electron identification down to 0.6 deg. Electron tagging at the lowest angles primarily serves to suppress four-fermion SM processes and other background processes with the characteristic low-angle electron signature.

## 4. Signal and background

### 4.1 Event samples

The SM-like Higgs boson with a mass of 125 GeV is dominantly produced via $W^+W^-$ fusion in $e^+e^-$ collisions at 1.4 TeV centre-of-mass energy. Four years of detector operation with a 50% data taking efficiency is assumed. This will lead to the total integrated luminosity of 1.5 ab$^{-1}$ at the nominal peak luminosity of $1.3 \cdot 10^{34}$ cm$^{-2}$s$^{-1}$. Unpolarised beams are considered. The Higgs production cross section above 1 TeV can be determined with a statistical precision better than 1% [1]. However, signal statistics is small due to the low branching fraction of Higgs to $\mu^+\mu^-$ decay being of order of $10^{-4}$. Higgs production through $W^+W^-$ fusion can be statistically enhanced by a factor 2.34 with the proper beam polarization [1].

Concurrent Higgs production via ZZ fusion is an order of magnitude lower. On a sample of 300 events of ZZ fusion followed by the Higgs decay to a pair of muons, not a single event passed the requirements applied in this analysis, implying a selection efficiency smaller than 1.2% (95% CL) for this channel.

In order to obtain accurate di-muon invariant mass distributions for signal and background, sufficiently large numbers of events are simulated. In Table 1, the full list of physics and beam-induced backgrounds is given.

Emission of beamstrahlung photons is the main cause for the low-energy tail in the luminosity spectrum. Beamstrahlung also produces incoherent $e^+e^-$ pairs deposited mainly at low-angle calorimeters as well as $\gamma\gamma \rightarrow \nu_\mu \bar{\nu}_\mu \mu^+ \mu^-$ with a very high cross-section of 110.7 fb. In addition, beamstrahlung photons produce hadrons ($\gamma\gamma \rightarrow$hadrons) as often as 1.3 interactions per bunch-crossing, influencing muon reconstruction efficiency due to the occupancy of the inner tracker. Deposited hadrons consequently produce tails in visible energy distribution of the signal. Processes $\gamma\gamma \rightarrow$hadrons and creation of incoherent pairs from beamstrahlung are included in the analysis by overlaying each event after full simulation and before the digitization phase.

### 4.2 Signal and background kinematic properties and MVA based selection

Based on the properties of signal and background, the reconstruction of two muons in an event, a di-muon invariant mass in the range 105-135 GeV and the absence of a high-energy electron (E>200 GeV) at one side of the very forward calorimeters serve as a preselection in this analysis. By vetoing electron-tagged events at the preselection stage, rejection rates for four-fermion and $e^\pm\gamma \to e^\pm\mu^+\mu^-$ processes are 25% and 15% respectively. Signal is rejected in only 0.2% of cases.

In addition, distributions of the following sensitive observables are combined in a multivariate approach to define optimal selection for various background processes:
- visible energy of the event $E_{vis}$;
- transverse momentum of the di-muon system $p_T(\mu\mu)$;
- scalar sum of the transverse momenta of the two selected muons $p_T(\mu_1)+p_T(\mu_2)$;
- relativistic velocity of the di-muon system $\beta(\mu\mu)$;
- polar angle of the di-muon system $\theta(\mu\mu)$;
- helicity angle $\cos\theta^*$, $\cos\theta^* = \dfrac{\vec{p}\,'(\mu_1) \cdot \vec{p}(\mu\mu)}{|\vec{p}\,'(\mu_1)| \cdot |\vec{p}(\mu\mu)|}$, where prime denotes rest frame of the di-muon system. It is sufficient to take only one muon ($\mu_1$) as the muons are back to back in the di-muon reference frame.

Distributions of these kinematic variables for $e^+e^- \to e^+e^-\mu^+\mu^-$, $\gamma\gamma \to \nu_\mu\bar{\nu}_\mu\mu^+\mu^-$ and the signal are used to train TMVA with the Boosted Decision Tree (BDT) classifier. The training is performed using 6000 events of signal, 250000 $e^+e^- \to e^+e^-\mu^+\mu^-$ events and 90000 $\gamma\gamma \to \nu_\mu\bar{\nu}_\mu\mu^+\mu^-$ events. Classifier output cut-off value (BDT>0.084) is selected to minimise relative statistical error of the measurement. Background from four-fermion processes is reduced by almost 3 orders of magnitude and the overall signal efficiency is 28%, including reconstruction, preselection and MVA selection. The impact of coincident tagging of Bhabha events is not considered.

Table 1. List of considered processes (1 signal and 2-9 background) with a corresponding cross-sections and number of simulated events. a) Including 100 GeV <$m(\mu^+\mu^-)$<140 GeV and a minimal polar angle of 8 deg for each muon. b) Photons originate from beamstrahlung.

| Process | $\sigma$[fb] | N |
|---|---|---|
| 1) $e^+e^- \to h\nu_e\bar{\nu}_e, H \to \mu^+\mu^-$ | 0.052 | 24000 |
| 2) $e^+e^- \to \mu^+\mu^-\nu_e\bar{\nu}_e$ | 129 | 236000 |
| 3) $e^+e^- \to e^+e^-\mu^+\mu^-$ | 431[a] | 1000000 |
| 4) $e^\pm\gamma \to e^\pm\mu^+\mu^-$ [b] | 1920[a] | 2000000 |
| 5) $\gamma\gamma \to \nu_\mu\bar{\nu}_\mu\mu^+\mu^-$ [b] | 110.7 | 350000 |
| 6) $e^+e^- \to \nu_\tau\bar{\nu}_\tau\tau^+\tau^-$ | 84.5 | 133000 |
| 7) $e^+e^- \to e^+e^-\tau^+\tau^-$ | 1942.2 | 464500 |
| 8) $e^+e^- \to \mu^+\mu^-$ | 17 | 50000 |
| 9) $e^+e^- \to \tau^+\tau^-$ | 358 | 482500 |

# 5. Di-muon invariant mass fit and $\sigma_{H\nu_e\bar{\nu}_e} \cdot BR(H\rightarrow\mu^+\mu^-)$ extraction

In order to determine branching fraction $BR(H\rightarrow\mu^+\mu^-)$, the number of selected signal events $N_s$ has to be known,

$$\sigma_{H\nu_e\bar{\nu}_e} \cdot BR(H\rightarrow\mu^+\mu^-) = N_s/(L \cdot \varepsilon_S) \qquad (1)$$

where $\sigma_{H\nu_e\bar{\nu}_e}$ stands for the Higgs production cross-section in $W^+W^-$ fusion, and L and $\varepsilon_S$ are the integrated luminosity and the signal selection efficiency, respectively. The number of signal events is determined by fitting the model of signal and background PDFs on simulated distribution of the invariant mass of the dimuon system.

In order to estimate the statistical uncertainty of the measurement and the fit, 5000 Toy Monte Carlo experiments were performed, where pseudo-data is obtained from randomly sampled fully simulated signal events and by random generation of the invariant mass from the background PDF. The sample sizes correspond to the common luminosity L of 1.5 ab$^{-1}$, selection efficiency $\varepsilon_i$ and the corresponding cross-section $\sigma_i$, ($N_i=\sigma_i \cdot \varepsilon_i \cdot L$). For each toy MC experiment, pseudo-data di-muon invariant mass distribution is fitted by function f:

$$f = k \cdot f_S + (1-k) \cdot f_{BCK} \qquad (2)$$

to determine the number of signal events as:

$$N_S = k \cdot \int f_S dm \qquad (3)$$

where $f_S$ and $f_{BCK}$ stand for signal and background PDFs, k is normalization coefficient and integration is performed in the mass region (100-140) GeV. Statistical uncertainty of the signal count is determined as RMS of the $N_s$ distribution of the all performed pseudo-experiments.

Fully simulated, as large as possible samples of signal and background (Table 1) are fitted to extract PDFs. Signal is fitted with a composite Gaussian function $f_S = t_1 + C_1 \cdot t_2$ with a flat and exponential tail, where the functions $t_{1,2}$ are defined in Figure 1. Total background (as given in Table 1) is fitted with a function containing flat and exponential part $f_{BCK} = p_0(p_1 e^{p_2(m-m_H)} + (1-p_1))$. Fits for signal and background, together with the fit parameters are given in Figure 1.

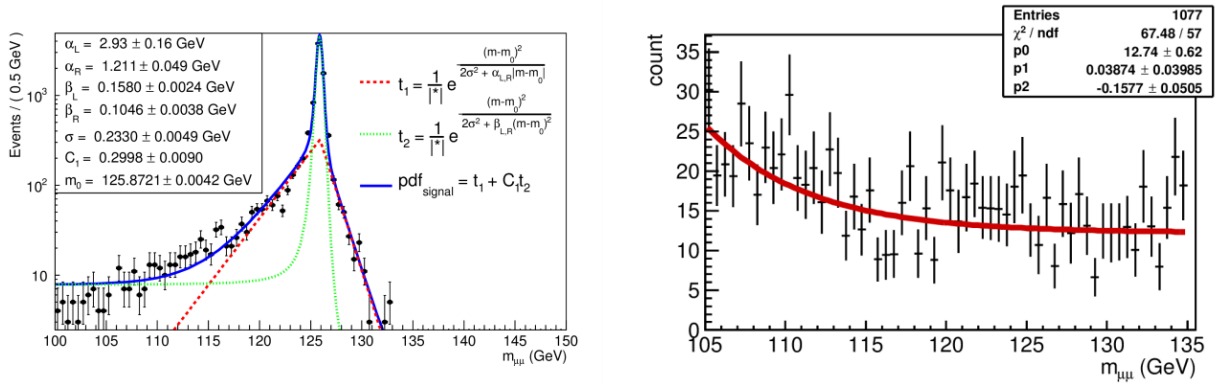

Figure 1. PDFs are determined from the fit of di-muon invariant mass distributions for signal (left) and background (right).

The overall function f (Eq. 2) is used to fit the expected shape of pseudo-data with an unbinned likelihood fit for each Toy Monte Carlo experiment. An example of one Toy Monte Carlo fit is given in Figure 2 (left). From Figure 2 it is clear that the statistics of the signal is limited to a few tens of events. The RMS of the number of signal events per experiment corresponds to the statistical error of the measurement (Figure 2 right).

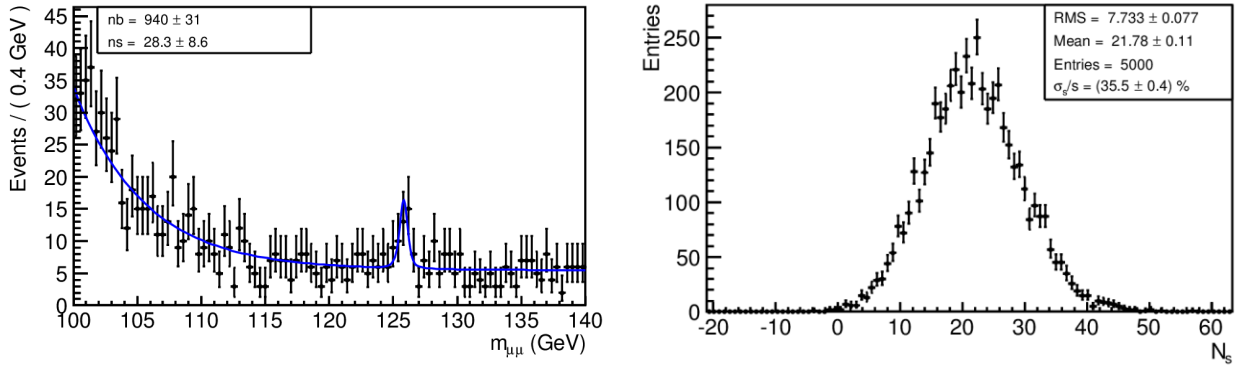

Figure 2. Di-muon invariant mass fit from a Toy Monte Carlo experiment (left). Number of signal events per Toy Monte Carlo experiment (right).

Uncertainty of the product $\sigma_{H\nu_e\bar{\nu}_e} \cdot BR(H\to\mu^+\mu^-)$ translates into the uncertainty of the observable $g^2_{HWW} \cdot g^2_{H\mu\mu}/\Gamma_H$, where g stands for a Higgs coupling and $\Gamma_H$ for the Higgs boson width. As can be seen from Figure 2 right, the relative statistical uncertainty of $\sigma_{H\nu_e\bar{\nu}_e} \cdot BR(H\to\mu^+\mu^-)$ is 35.5%. Since the Higgs coupling to EW bosons can be determined at CLIC to within a few percent ($\delta(g_{HWW})$=2.1%), as well as the Higgs width ($\delta\Gamma_{H, model}$)=0.3%) [1], the corresponding uncertainty of the Higgs coupling to muons $g_{H\mu\mu}$ is estimated to be 15.4%. The measurement is dominated by limited signal statistics and the presence of irreducible backgrounds primarily from $e^+e^- \to \mu^+\mu^-\nu_e\bar{\nu}_e$ processes. Systematic uncertainties originating from the uncertainty of the peak-luminosity, uncertainty on momentum resolution and uncertainty of the muon identification efficiency result in a percent order effect. Accidental signal rejection due to coincident Bhabha electron tagging is not yet considered.

Results presented in these proceedings are quantitatively different from the once presented at the LCWS13 due to the fact that $\gamma\gamma \to \nu_\mu\bar{\nu}_\mu\mu^+\mu^-$ processes are now included in the analysis at the full simulation level.

## 6. Conclusions

There is a strong motivation to search for signs of physics beyond the Standard Model at CLIC. Measurements of Higgs boson couplings are of particular interest.

The measurement of the branching ratio for the rare SM-like Higgs decay into two muons is simulated at 1.4 TeV CLIC with unpolarised beams. The measurement itself tests the muon identification and momentum resolution of the detector.

It was shown that measurement of the branching ratio for the Standard Model Higgs decay into two muons can be performed with a statistical uncertainty of ~35.5% and a systematic uncertainty at the percent level. The largest contributions to the uncertainty of the measurement are limited statistics of the signal and the presence of signal-like backgrounds. Statistical uncertainty of the BR measurement translates into precision of the Higgs coupling to muons $g_{H\mu\mu}$ of 15.4%.

EPA approximation in the cross-section generation of appropriate background processes as well as the impact of signal rejection due to coincidental tagging of Bhabha electron in the very forward region will be introduced to the analysis as the next steps.

## 7. References

[1] H. Abramowicz, *et al.*, Physics at the CLIC $e^+e^-$ Linear Collider - Input to the Snowmass process 2013, July 2013, arXiv:1307.5288
[2] M. A. Thomson, *et al.*, The physics benchmark processes for the detector performance studies of the CLIC CDR, 2011, CERN LCD-Note-2011-016.
[3] W. Kilian, T. Ohl, and J. Reuter. WHIZARD: Simulating multi-particle processes at LHC and ILC. 2007, arXiv:0708.4233v1
[4] T. Sjostrand, S. Mrenna, and P. Z. Skands. PYTHIA 6.4 Physics and Manual. JHEP, vol. 05 p. 026, 2006, hep-ph/0603175.
[5] Z. Was. TAUOLA the library for tau lepton decay, and KKMC/KORALB/KORALZ/... status report. Nucl. Phys. Proc. Suppl., vol. 98 pp. 96–102, 2001, hep-ph/0011305.
[6] D. Schulte. Beam-beam simulations with GUINEA-PIG. 1999. CERN-PS-99-014-LP.
[7] A. Münnich and A. Sailer, The CLIC_ILD_CDR geometry for for the CDR Monte Carlo mass production, 2011, CERN LCD-Note-2011-002.
[8] P. Mora de Freitas and H. Videau, Detector simulation with MOKKA / GEANT4: Present and future, prepared for International Workshop on Linear Colliders (LCWS 2002), Jeju Island, Korea, 26-30 August 2002.
[9] M. A. Thomson. Particle Flow Calorimetry and the PandoraPFA Algorithm. Nucl. Instr. Meth., vol. A611 pp. 25–40, 2009, arXiv:0907.3577.
[10] A. Hocker, *et al.*, TMVA - Toolkit for multivariate data analysis 2009, arXiv:physics/0703039
[11] Physics and Detectors at CLIC: CLIC Conceptual Design Report, 2012, arXiv:1202.5940